\documentclass[prl, floatfix,aps,showpacs,twocolumn,nofootinbib,reprints, preprintnumbers]
{revtex4}
\usepackage{graphicx}

\begin{document}

\preprint{}

\title{Out-of-equilibrium Baryogenesis and SuperWIMP Dark Matter}

\author{Giorgio Arcadi}
\affiliation{Institut f\"ur theoretische Physik, Friedrich-Hund-Platz 1, D-37077 G\"ottingen, Germany}
\author{Laura Covi}
\affiliation{Institut f\"ur theoretische Physik, Friedrich-Hund-Platz 1, D-37077 G\"ottingen, Germany}
\author{Marco Nardecchia}
\affiliation{DAMTP, CMS, University of Cambridge, Wilberforce Road, Cambridge CB3 0WA,
United Kingdom\\
Cavendish Laboratory, University of Cambridge, JJ Thomson Avenue, Cambridge
CB3 0HE, United Kingdom }

\begin{abstract}
The experimental fact that the energy density in Dark Matter and in Baryons is of the
same order is one of the most puzzling in cosmology. In this letter we suggest a new
mechanism able to explain this coincidence in the context of out-of-equilibrium 
baryogenesis  with DM production "{\it \'a la}" SuperWIMP starting from the same initial particle. 
We then discuss two simple implementations of this scenario within supersymmetric
models with gravitino DM. 
\end{abstract}

\pacs{98.80.Cq}

\maketitle

\section{Introduction}

It is an amazing coincidence that both Dark Matter and the baryonic energy densities are 
approximately within a factor five of each other \cite{Ade:2013zuv}. Since normally the two densities are 
generated by very different mechanisms and at different scales, e.g. a WIMP mechanism below 
the electroweak scale and leptogenesis at high scale, it is usually not possible to understand 
theoretically why the numbers are not much wider apart.
Since the pioneering work by  Sakharov~\cite{Sakharov}, it has been realized that obtaining a sufficiently 
large baryon number is usually a much more difficult task than just to produce Dark Matter: 
baryogenesis requires a sufficiently large violation of C, CP and baryon number violation and a 
departure from thermal equilibrium, while DM production can take place even in thermal equilibrium 
and without any quantum number violation.
One would therefore expect the density of Dark Matter to be much less suppressed than the
baryon asymmetry and indeed, in order to obtain the observed number density for Dark
Matter, one usually has to rely on either a reduced number density, e.g. via a non-relativistic
 decoupling, or a very small mass for the Dark Matter particle, which is often in tension with
 being {\it Cold} Dark Matter. 

A simple way to connect Dark and Baryonic Matter is to invoke for both types of matter an
asymmetry, like it has been proposed in Asymmetric Dark Matter (ADM) models \cite{Zurek:2013wia}.
In that case, the asymmetries in the two species can be related, and then in the simplest realizations
the ratio between the matter densities can be simply explained through a ratio of masses.
Then it is expected that the Dark Matter has a mass not much heavier than the proton and must 
interact sufficiently strongly to erase the symmetric density component.

In a similar spirit, in this letter we would like to propose and explore another way to connect the baryon and DM
generation relying on baryogenesis through out-of equilibrium decay~\cite{Kolb:1979qa, GUT-baryogenesis} 
and the SuperWIMP mechanism~\cite{Covi:1999ty,Feng:2003xh,Feng:2003uy} . 
In such a case both matter densities are produced from an initial mother particle and 
they are naturally suppressed compared to its initial density, by the CP violation, needed to 
generate the baryon asymmetry,  and by the branching ratio into DM respectively. 
These two suppression factors can be naturally of the same order of magnitude and explain 
why the baryon and Dark Matter densities turn out to be so similar. 
In general then the ratio between  the DM and the baryonic energy densities is independent 
of the original mother particle density and given just by masses, the decay CP asymmetry 
and branching ratios.

\section{The basic mechanism}

The mechanism we would like to propose is very simple and relies on the possibility of generating both
the baryon asymmetry and  Dark Matter via the out-of-equilibrium decay of the same particle~$X$.
Let us consider first baryogenesis. In general if a massive particle $X$ decays out-of-equilibrium 
in two channels with different baryon number and with a non-vanishing C- and
CP-violation, one can obtain from the decay the baryon number~\cite{Kolb:1979qa, GUT-baryogenesis}
\begin{equation}
\Omega_{b} = \xi_b \epsilon_{\rm CP} \frac{m_p}{m_{X}}\; BR(X \rightarrow b, \bar b)\;  \Omega_{X}
\end{equation}
where $\xi_b $ is a coefficient taking into account the possible effects of wash-out processes
and baryon number dilution (e.g. via annihilation of massive particles into photons),
$m_p, m_X $ are the proton and decaying particle masses,  $\Omega_{X} $ is the initial density of the 
$X$ particle at departure of equilibrium and $BR(X \rightarrow b, \bar b) $ gives the branching ratio 
for the decay into baryons and antibaryons.
The CP-violation in the decay is taken into account by $\epsilon_{CP}$ given as:
\begin{eqnarray}
\epsilon_{\rm CP} &=& 
\frac{\Gamma (X \rightarrow b) - \Gamma (X \rightarrow \bar b)}{\Gamma (X \rightarrow b) +\Gamma (X \rightarrow \bar b)}\; .
\end{eqnarray}
In order to generate a sufficiently large $\Omega_b $ one needs a large initial number density 
of the $X$-particle since both $\xi $ and $\epsilon_{CP} $ tend to suppress the final baryon density.
For the general case of WIMP-like decoupling, $\Omega_{X} \propto m_{X}^2 $, this condition requires 
a heavy $X$-particle possibly above the TeV scale with suppressed annihilation channels.

In this setting, we consider also the decay of the $X$-particle into Dark Matter.
The presence of this additional decay channel does not modify the mechanism of  baryogenesis discussed above, 
as long as the branching ratio into DM is negligibly small. The decays of the particle $X$ after its freeze-out 
produce a DM abundance through the 
SuperWIMP mechanism~\cite{Covi:1999ty,Feng:2003xh,Feng:2003uy,Arcadi:2013aba} as:
\begin{equation}
\Omega_{DM} = \xi_{DM} \frac{m_{DM}}{m_{X}} BR\left(X\rightarrow DM+\mbox{anything} \right) \Omega_{X}
\end{equation} 
where in this case $ \xi_{DM} $ just accounts for the possible dilution after DM production.
Therefore due to the presence of the other decay channels, the DM density is suppressed by the 
corresponding branching ratio and is much smaller than the original $X$ density.
Note that the time-scale of the $X$-particle decay is set by the total decay rate $\Gamma_{tot} $ avoiding 
problems with Big Bang Nucleosynthesis as long as $\Gamma_{tot}^{-1} < 1 $ s.

We see that in this scenario we expect both $\Omega_b $ and $\Omega_{DM} $ to be suppressed
by small numbers compared to $\Omega_{X} $ and we can obtain their ratio as
\begin{eqnarray}
\label{eq:ratio_fit}
\frac{\Omega_b}{\Omega_{DM}}\!\! &=&\!\! \xi\,\epsilon_{\rm CP} \frac{m_p}{m_{DM}} 
\frac{BR(X \rightarrow b, \bar b)}{BR\left(X \rightarrow DM +\mbox{anything} \right)}\; ,
\end{eqnarray}
with $\xi = \xi_b/\xi_{DM} $, independently of the initial density of the particle $X$. So we can indeed obtain naturally 
$\Omega_{b}/\Omega_{\rm DM} \sim 1/5$, if the masses of the DM and of the proton are 
of the same order and if the branching ratio of the decay of $X$ into DM is strongly suppressed 
in comparison to the other channels and is of order $\epsilon_{\rm CP} $.
Note that in order to explain the whole Dark Matter abundance as coming from the same particle that
produces the baryon asymmetry, all the other mechanisms that could produce DM  in the early universe have to be subdominant. 

In the following we will discuss two different implementations of this mechanism in the context of supersymmetric
models with gravitino DM. Indeed if the $X$ particle is a superpartner, it has always a decay channel to gravitino LSP 
with Planck suppressed decay rate. We have indeed for a fermionic $X$:
\begin{eqnarray}
\Gamma \left(X \rightarrow \tilde{G}+\mbox{anything}\right) &=& \frac{1}{48 \pi} \frac{m_X^5}{m_{3/2}^2 M_{\rm Pl}^2}
\end{eqnarray}
where  $M_{\rm Pl}$  is the reduced Planck mass, $M_{\rm Pl}=2.43 \times 10^{18}$GeV,
$M_X$ is the mother particle mass and $m_{3/2} $ the gravitino mass. 
It is in this case therefore easy to achieve a small branching ratio into gravitino of the order of $\epsilon_{\rm CP}$.
Indeed the condition (\ref{eq:ratio_fit}) allows for a gravitino lighter than the proton, but the requirement of proton
stability against processes mediated by the $X$-particle, points towards a gravitino heavier than 1 GeV, corresponding
to $ \epsilon_{\rm CP} \geq BR\left(X \rightarrow \tilde{G}+\mbox{anything} \right) $.

\section{A simple implementation of the mechanism at the TeV scale}

Let us consider one this mechanism at the TeV scale in the case when baryogenesis occurs in a supersymmetric model with R-parity 
violation~\cite{Cui:2012jh,Cui:2013bta,Sorbello:2013xwa}. The model is based on the MSSM extended with the baryon and R-parity
violating superpotential
\begin{equation}
W_{B\!\!\!/} = \lambda_{ijk}'' U^c_i D^c_j D^c_k
\end{equation}
where $ U^c, D^c $ represent the squark chiral multiplets and $i,j,k$ are generation indices. 
Then a Bino-like neutralino, not the LSP, can decay into different R-parity conserving channels, in particular the
dominant one into a gluino and a quark-antiquark pair and subdominant decays into photon/Z and gravitino, as well as 
into three quarks via the R-parity violating coupling. Then the Bino plays the role of the $X$ particle with decay 
rates given by~\cite{Baltz:1997gd, Cui:2013bta}
\begin{eqnarray}
\Gamma\left(\tilde B \rightarrow \tilde{g} q_i \overline{q_i} \right) &=& \frac{Y_i^2 \alpha_1 \alpha_3 }{16 \pi} \frac{m_{\tilde B}^5}{m_0^4} \cr
\Gamma\left(\tilde B \rightarrow u_i d_j d_k + \bar u_i\bar d_j \bar d_k \right) &=& \frac{ |\lambda_{ijk}'' |^2 \alpha_1 (Y_u-Y_d)^2}{16 \pi^2} \frac{m_{\tilde B}^5}{m_0^4} \cr
\Gamma \left(\tilde B \rightarrow \tilde{G}+\mbox{anything}\right) &=& \frac{1}{48 \pi} \frac{m_{\tilde B}^5}{m_{3/2}^2 M_{\rm Pl}^2}
\end{eqnarray}
where  $\alpha_1, \alpha_3 $ are the gauge coupling strenghts of the $U(1)_Y$ and $SU(3)$ gauge groups, 
$Y_{u,d,q} = 2/3, -1/3, 1/6 $ are the hypercharges 
of the quarks $u_R, d_R, q_L$, $m_{\tilde B}$ and $m_0$ denote the Bino's and a common scalar superpartner's masses.
In the rate we have already summed over color states, while the sum over the final state flavours gives factors 
$ N_{RPC} = 6 Y_q^2 + 3 Y_u^2 + 3 Y_d^2 =  11/6 $ instead of $ Y_i^2 $ in the R-parity conserving case and 
$ N_{RPV} = 9 (Y_u-Y_d)^2 = 9$ instead of  $(Y_u-Y_d)^2$ for the RPV decay for flavour-universal $\lambda''$. 
If some of the squarks are substantially heavier than the others, the rates get smaller and we can obtain the limiting 
cases with decoupled $ \tilde u_R, \tilde d_R, \tilde q_L $ by setting $ Y_u, Y_d, Y_q $ equal to zero respectively.
In order for the neutralino to decay out-of-equilibrium, the scalar superpartners have to be sufficiently heavy 
to satisfy the relation
$\Gamma_{tot} < H(T=m_{\tilde B}) $
but lighter than the SUSY breaking scale $ \sqrt{m_{3/2} M_{Pl}} $ in order for the decay into gravitino to have a small branching ratio.
For universal $ \lambda ''$, we obtain from these two conditions the window of squark mass:
\begin{equation}
\sqrt{m_{3/2} M_{Pl}} \gg (3 A)^{-1/4} m_0 > \frac{1}{2\sqrt{\pi}} \left(\frac{g_\star}{10} \right)^{-1/8} m_{\tilde B}^{3/4} M_{Pl}^{1/4} 
\end{equation}
with $ A= \alpha_1 ( |\lambda''|^2/\pi N_{\rm RPV} + \alpha_3 N_{\rm RPC}) $ and $ g_\star $ counting the effective relativistic
degrees of freedom at the time when the neutralino becomes non-relativistic.

\begin{figure}
\includegraphics[width=2.8in]{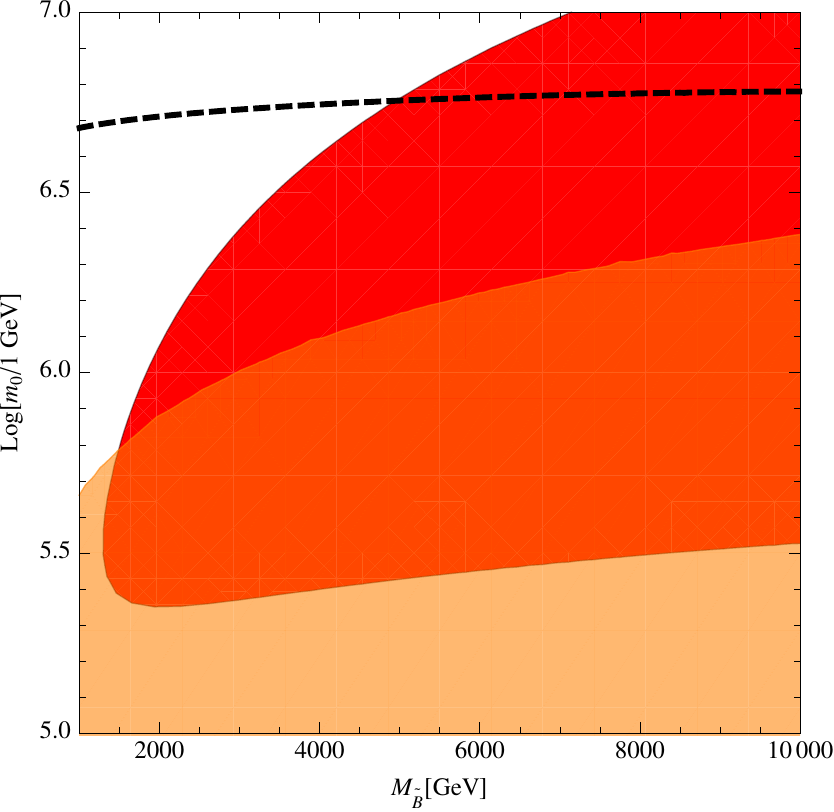}
\caption{\label{Fig1}  We show here in red (dark grey) the region where 
$\Omega_b $ is between 0.01-0.04 in the plane $m_0$ vs $ m_{\tilde B} $ for
$\lambda'' = 0.5 $. The black line gives $ \Omega_{DM} h^2 = 0.11$ for $ m_{3/2} = 1$ GeV.
We can see that we can obtain both quantities in the observed range for 
$m_0 \sim 5-6 \times 10^6 $ GeV and $ m_{\tilde B} \geq 6 $ TeV.
In the orange (light grey) region, the Bino annihilation and other wash-out processes are
still active during the decay.
}
\end{figure}

The RPV decay may have a non-vanishing CP asymmetry, due to the one-loop diagrams with on-shell gluino and quark. 
Assuming that the couplings contain different complex phases and that the intermediate down squarks dominate 
the process, the CP violation parameter can be estimated as~\cite{Cui:2013bta,acn14}
\begin{equation}
\epsilon_{\rm CP}=\frac{1}{20}Im\left[e^{i \phi}\right] 
\alpha_3  \frac{m_{\tilde{B}}^2}{m_0^2}
\end{equation}
where $ \phi $ denotes an effective CP-violating phase.
If the Bino density is sufficiently large, as it can be arranged for large $ \mu $ and pure Bino, i.e.
when the annihilation is mostly into Higgs bosons, a baryon asymmetry can be generated as~\cite{Cui:2013bta}
\begin{equation}
\Omega_b \sim 10^{-2}  \left(\frac{m_{\tilde{B}}}{1 \mbox{TeV}}\right)\!\!\! {\left(\frac{\mu}{10^{3/2} m_0}\right)}^2\!\! \left(\frac{\alpha_1 |\lambda''|^2 N_{\rm RPV}}{\pi\, A}\right)
\end{equation}
for $ Im\left[e^{i \phi}\right] = 1 $.

So we can obtain for the baryon to DM ratio simply
\begin{eqnarray}
\label{eq:ratio_fit2}
\!\!\frac{\Omega_b}{\Omega_{DM}}\!\!\!\! &\sim& \!\!
\xi N_{\rm RPV}  \left(\frac{m_{3/2}}{1 \mbox{GeV}}\right)\!\!
{\left(\frac{\lambda''}{0.1}\right)}^2\! {\left(\frac{m_{\tilde B}}{1 \mbox{TeV}}\right)}^2\! {\left(\frac{m_0}{10^3 \mbox{TeV}}\right)}^{-6} .
\nonumber\\
\end{eqnarray}
From this expression it is clear that we need $m_0 \gtrsim 10^3\mbox{TeV}$ to match the observed baryon to DM ratio and
that the gravitino mass cannot be far from the proton mass and, likely, values lower than 1 GeV are favored. 
Such a large scalar mass is also required to obtain the right baryon number \cite{Cui:2013bta,acn14}. 
We show in Figure 1 the regions of parameter space with the correct baryon and DM densities in the plane 
of $m_0$ vs $ m_{\tilde B} $ for a fixed gravitino mass of $1$ GeV.
In Figure 2 we give the same regions in the plane $m_{\tilde B} $ vs $ m_{3/2} $ for fixed values of $m_0 $ and $\mu$.

\begin{figure}
\includegraphics[width=2.8in]{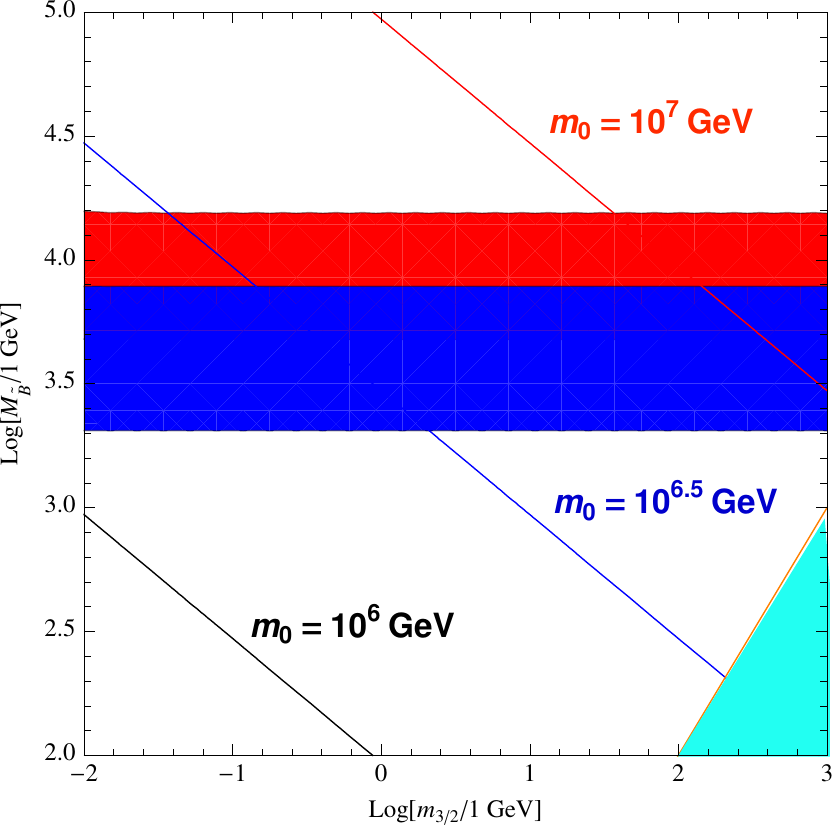}
\caption{\label{Fig2} We show here in the plane $m_{\tilde B} $ vs $ m_{3/2} $ 
the lines where the ratio $\Omega_b/\Omega_{DM} $ takes the observed value for different 
values of the scalar mass $m_0$. 
The matching colored regions have also the observed value for the baryon number,
assuming $\mu = 10^2 m_0 $. 
}
\end{figure}

In order for this mechanism to work, one has  to be sure that the contributions to the gravitino density from thermal scatterings \cite{Bolz:2000fu}
and FIMP \cite{Hall:2009bx,Cheung:2011nn} mechanism from the heavy scalars are not larger than the SuperWIMP one. The easiest way to achieve 
that simultaneously is to require a sufficiently low reheat temperature that the scalar superpartners are not in thermal equilibrium 
$ T_{RH} < m_0 $, while the FIMP decay of the gauginos is less important. A detailed study of all the relevant processes will
be the subject of a longer publication \cite{acn14}.

\section{A simple implementation of the mechanism at a high scale}

Let us consider now another implementation of the mechanism, where instead baryogenesis happens
via leptogenesis at a high scale \cite{Fukugita:1986hr}. In this case the decay of the lightest RH neutrino and its superpartner, the
RH sneutrino, both produce a lepton asymmetry with similar values of $\epsilon_{CP} $ \cite{Covi:1996wh}.
The lightest RH neutrino and sneutrino decay rates into leptons and antileptons and their superpartners are given by
\begin{eqnarray}
\Gamma_{L\!\!\!/, \tilde L\!\!\!/} = \Gamma\left(N, \tilde N \rightarrow L + \bar L \right) &=& 
\frac{(\lambda^\dagger \lambda)_{11} M_{N, \tilde N} }{4 \pi} 
\label{SNdecayLH}
\end{eqnarray}
so that to have an out-of-equilibrium decay we need
\begin{equation}
M_{N,\tilde N} > \frac{3}{4\pi^2} \sqrt{\frac{10}{g_\star}} (\lambda^\dagger \lambda)_{11} M_{Pl} 
\geq 10^{9} \mbox{GeV}.
\end{equation}
Note that barring cancellations the heaviest light neutrino mass is well approximate by 
$ \tilde m_1 \sim (\lambda^\dagger \lambda)_{11} v_u^2/M_N \leq 1\; \mbox{eV} $ \cite{Buchmuller:2004nz}.
The generated lepton asymmetry is given by
\begin{equation}
Y_{\ell} = \xi_l \left(\epsilon_{CP, N} Y_N + \epsilon_{CP,\tilde N} Y_{\tilde N} \right) \; ,
\end{equation}
where again $\xi_l $ takes into account possible effects of wash-out processes as well as
the change in degrees of freedom from the leptogenesis epoch to today and $ Y_i = n_i/s $
is the number density of the species $i$ rescaled by the entropy density $s$.
In the preferred region of strong wash-out, where the baryon asymmetry is independent of the initial conditions
\cite{Buchmuller:2004nz}, we have $ \xi_l \sim 10^{-4}-10^{-5} $.
This lepton asymmetry is finally transferred into a baryon asymmetry by sphaleron processes as
\begin{equation}
\Omega_b = c_{SF} m_p Y_{\ell} 
\end{equation}
where  $ c_{SF} =  8/23 $ is the sphaleron factor relating the original $B-L$ number to the baryon number \cite{Harvey:1990qw}.

In this setting the production of gravitino Dark Matter can happen via RH sneutrino decay. If the degeneracy in mass between the 
RH sneutrino and neutrino is lifted, i.e. for  $\Delta M_N^2 \equiv  M_{\tilde N}^2 - M_{N}^2 > 0$ and 
$ M_{\tilde N} - M_{N} \simeq \frac{\Delta M_N^2}{2 M_N} > m_{3/2} $
 the decay can proceed into RH neutrino and gravitino.  This points towards large mass splitting
$\Delta M_N^2 $.
The decay rate of the RH sneutrino into a RH neutrino and a gravitino is
\begin{eqnarray}
\Gamma \left(\tilde N \rightarrow \tilde{G} + N\right) &=& \frac{1}{48 \pi} \frac{M_{\tilde N}^5}{m_{3/2}^2 M_{\rm Pl}^2}
\left(1- \frac{M_{N}^2}{M_{\tilde N}^2}  \right)^4 \cr
&\sim& \frac{1}{48 \pi} \frac{\Delta M_N^8}{m_{3/2}^2 M_{\rm Pl}^2 M_{\tilde N}^3}
\label{SNdecay2}
\end{eqnarray}
so that the decay is suppressed both by $M_{Pl}$ and the available phase space. 
Indeed the decay is governed by the Goldstino coupling that vanishes in the limit
of conserved SUSY. 
If the mass splitting is too small, 
the three-body decay into Higgs, lepton and gravitino can be more important.
For such three-body decay, away from resonances in the intermediate RH neutrino or higgsino, 
taken into account in Eqs.~(\ref{SNdecayLH}), (\ref{SNdecay2}),
we have
\begin{eqnarray}
\Gamma \left(\tilde N \rightarrow \tilde{G} + \ell h\right) &=& 
\frac{(\lambda^\dagger \lambda)_{11}}{1536 \pi^3} \frac{M_{\tilde N} \Delta M_N^4}{m_{3/2}^2 M_{\rm Pl}^2} \cr
&\sim& 
\frac{\Gamma_{\tilde L\!\!\!/}}{384 \pi^2} \frac{\Delta M_N^4}{m_{3/2}^2 M_{\rm Pl}^2}.
\end{eqnarray}
We see therefore that the 3-body decay dominates over the 2-body as soon as
\begin{equation}
\frac{\Delta M_N^4}{M_{\tilde N}^4} < \frac{(\lambda^\dagger \lambda)_{11}}{32 \pi^2}.
\end{equation}

The branching ratio of the RH sneutrino decay into gravitino is then
\begin{equation}
BR (\tilde N \rightarrow \tilde G ) = \frac{\Delta M_N^4}{384 \pi^2 m_{3/2}^2 M_{\rm Pl}^2} 
\left( 1 +  \frac{\Delta M_N^4}{M_{\tilde N}^4} \frac{32\pi^2}{(\lambda^\dagger \lambda)_{11}}\right) 
\end{equation}
and it can naturally be of the same order as $ \epsilon_{CP}  \simeq 10^{-6} $ even for small $\lambda $.
Taking $ 0 \leq Y_{N} \sim Y_{\tilde N} $ and $\epsilon_{CP,N} \sim \epsilon_{CP,\tilde N} = \epsilon_{\rm CP}$, 
we have then for the ratio between baryon and DM energy density the relation:
\begin{eqnarray}
\frac{\Omega_b}{\Omega_{DM}} &=&   768 \pi^2 c_{SF} \xi \epsilon_{\rm CP}  \frac{m_p m_{3/2} M_{\rm Pl}^2}{\Delta M_N^4} \cr
& & \times \left( 1 +  \frac{\Delta M_N^4}{M_{\tilde N}^4} \frac{32\pi^2}{(\lambda^\dagger \lambda)_{11}}\right)^{-1}.
\end{eqnarray}
Assuming for the parameters the typical values from thermal leptogenesis~\cite{Buchmuller:2004nz}, 
i.e. $ M_{\tilde N} \sim 10^{10} $ GeV, 
$ (\lambda^\dagger \lambda)_{11} \sim 10^{-5} $, $ \epsilon_{\rm CP} \sim 10^{-6} $, and the strong 
wash-out regime with $ c_{sf} \xi \sim 10^{-3} $, 
we obtain the observed value of the ratio for intermediate values of the mass 
difference $\Delta M_N^2 $, i.e.
\begin{equation}
\sqrt{\Delta M_N^2} \sim 
1.2\times 10^{8} \, \mbox{GeV}  \left( \frac{m_{3/2}}{1\; \mbox{GeV}} \right)^{1/4}.
\end{equation}
Note that the result is independent of the RH sneutrino mass since the 3-body decay into gravitino 
dominates and has the same dependence on that mass as the lepton violating decays.
For larger gravitino masses instead the 2-body decay plays the dominant  role and the relation changes to
\begin{equation}
\sqrt{\Delta M_N^2} \sim 
3.3 \times 10^{8} \mbox{GeV}  \left( \frac{m_{3/2}}{1\; \mbox{TeV}} \right)^{1/8}  \left( \frac{M_{\tilde N}}{10^{10}\; \mbox{GeV}} \right)^{1/2} 
\end{equation}
for $ (\lambda^\dagger \lambda)_{11} \sim 10^{-5} $. 
We show in Figure 3 the curves for the observed value of $ \Omega_b/\Omega_{DM}$
in the plane $\epsilon_{CP} $ vs gravitino mass. For values $ \sqrt{\Delta M_N^2} \sim 10^7-10^8 $ GeV
it is possible to match the observation for $\epsilon_{CP} \sim 10^{-6} $ as required by thermal leptogenesis.
In case of non-thermal leptogenesis, the RH sneutrino and neutrino number densities
can be larger and allow also for smaller values of $ \epsilon_{CP} $ and therefore $ \Delta M_N^2 $.
\begin{figure}
\includegraphics[width=2.8in]{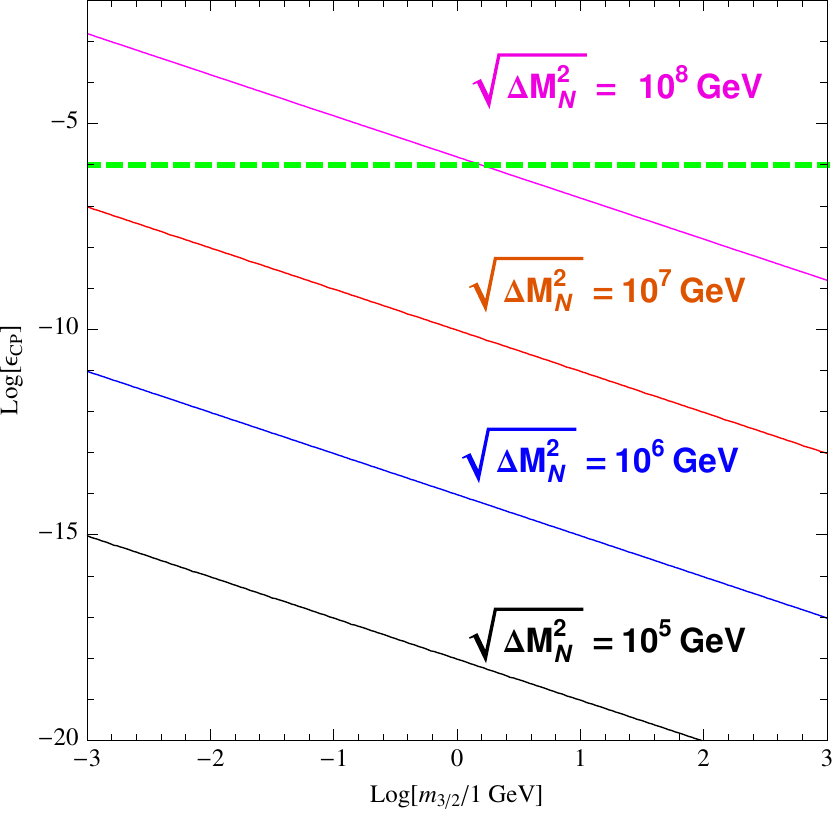}
\caption{\label{Fig3} Curves of correct values of the ratio $\Omega_b/\Omega_{DM} $ in the plane
$ \epsilon $ vs $ m_{3/2} $ for different values of the mass difference $ \Delta M_N^2 $.
We take $ M_{\tilde N} = M_N = 10^{10} $ GeV and $ (\lambda^\dagger\lambda )_{11} \sim 10^{-5}$.
A typical value $\epsilon_{CP} = 10^{-6} $ is given by the green (dashed) line. }
\end{figure}

In both cases we need a quite large mass splitting in the RH neutrino multiplet, larger than
expected for the other SM superpartners, but still smaller than the RH sneutrino masses.
Such a splitting could arise naturally  if the RH sneutrino couples directly to the supersymmetry breaking sector.

In this case again, in order for the mechanism to work, we need to suppress all other production channels
of gravitinos, which is not so simple at high temperature. Possibly the easiest way would be to consider a heavy 
gravitino that couples more weakly or to reduce the scale of leptogenesis and so also $ M_{\tilde N} $ as in 
non-thermal leptogenesis \cite{Giudice:1999fb, Asaka:2002zu}.

\section{Conclusions}

We have considered the possibility of producing both DM and the baryon asymmetry from the out-of-equilibrium decay of a single
particle. In the general case this allows to obtain densities of the same order since both the baryon and the DM densities are
strongly suppressed, by the CP asymmetry or by the decay branching ratio respectively. We are in this way able to obtain the observed ratio
just from fundamental parameters like masses and couplings, independently of the mother particle density.

The mechanism can in principle be embedded both in low and high scale baryogenesis models with gravitino DM, as we have discussed 
in sections 3 and 4, as long as the usual thermal gravitino production by scatterings or decays is sufficiently suppressed.
In both the models we find that at least part of the supersymmetric spectrum must be quite heavy. In the first case the squarks are required to be 
much heavier than the gauginos in order to achieve the out-of-equilibrium condition and also the right branching ratio of the Bino into gravitino. 
In the case of leptogenesis, a sufficiently large mass splitting in the RH neutrino multiplet is needed, implying very heavy scalar superpartners
if extended to the whole MSSM superpartners. Nevertheless the gravitino LSP and also other part of the spectrum can still reside below
or at the TeV scale.

\acknowledgments
We would like to thank Yossi Nir, Riccardo Rattazzi, Fabrizio Rompineve Sorbello and Damien George  for 
useful discussions.
G.A. thanks the CERN Theory Division and the LPTH Orsay for the warm
hospitality during part of the completion of this work.

G.A. and L.C. acknowledge partial financial support  by the EU FP7 ITN Invisibles 
(Marie Curie Actions, PITN-GA-2011-289442). L.C. acknowledges partial financial support 
by the German-Israeli Foundation for scientific research and development (GIF) .



\end{document}